\documentclass[a4paper,11pt]{article}

\usepackage{jinstpub} 

\title{Prototype Analog Front-end for Negative-ion Gas and Dual-phase Liquid-Ar TPCs}

\author[a]{Miki Nakazawa,}
\author[b,1]{Tetsuichi Kishishita,\note{Corresponding author.}}
\author[b]{Masayoshi Shoji, Ken Sakashita,}
\author[a]{Tomonori Ikeda,Hirohisa Ishiura}
\author[c]{James B. R. Battat, Catherine Nicoloff,}
\author[b]{Manobu M. Tanaka, Takuya Hasegawa,}
\author[a]{Kentaro Miuchi}

\affiliation[a]{Kobe University, 657-0013, 1-1 Nada-ku Rokkodaicho Kobe Hyogo, Japan}
\affiliation[b]{KEK High Energy Accelerator Research Organization, 305-0801, 1-1 Oho Tsukuba Ibaraki, Japan}
\affiliation[c]{Wellesley College Department of Physics, 106 Central Street, Wellesley, MA 02481, USA}

\emailAdd{kisisita@post.kek.jp}

\abstract{We report on the recent development of a versatile analog front-end compatible with  
 a negative-ion \mutpc{} for a directional dark matter search
 as well as a dual-phase, next-generation $\mathcal{O}$(10~kt) liquid argon TPC to study neutrino oscillations, nucleon decay, and astrophysical neutrinos.
Although the operating conditions for negative-ion and liquid argon TPCs are quite different (room temperature \vs{} $\sim$88~K operation, respectively), the readout electronics requirements are similar. Both require a wide-dynamic range up to 1600 fC, and less than 2000--5000 e$^-$ noise for a typical signal of 80 fC with a detector capacitance of $C_{\rm det} \approx 300$~pF. In order to fulfill such challenging requirements, a prototype ASIC was newly designed using 180-nm CMOS technology. Here, we report on the performance of this ASIC, including measurements of shaping time, dynamic range, and equivalent noise charge (ENC). We also demonstrate the first operation of this ASIC on a low-pressure negative-ion \mutpc{}.
}

\keywords{Front-end electronics for detector readout, Gaseous imaging and tracking detectors}

\newcommand{\vs}{\textit{vs.}}
\newcommand{\ie}{\textit{i.e.}}
\newcommand{\eg}{\textit{e.g.}}
\newcommand{\mutpc}{$\mu$-TPC}

\begin{document}
\maketitle
\flushbottom

\section{Introduction}
We report on a custom analog front-end electronics system developed by the KEK-Kobe University joint project {\it LTARS} (Low-Temperature Analog Readout System).
The project goal was to establish an analog readout system for the {\it NEWAGE} directional dark matter experiment (NEw general Wimp search with an Advanced Gaseous tracker Experiment) \cite{nakamura}, and for a next-generation neutrino oscillation experiment \cite{nuexp}.

The {\it NEWAGE} detector is a $\mu$-TPC (micro-Time Projection Chamber), which uses an MPGD (Micro Patterned Gaseous Detector), \eg{}, $\mu$-PIC \cite{miuchi} or Micromegas (Micro-MEsh Gaseous Structure) \cite{giomataris} to read out the TPC. {\it NEWAGE} aims to detect the directional signature of galactic dark matter \cite{tanimori, mayet}. Typically, TPCs operate in electron-drift mode: a charged nuclear recoil ionizes the surrounding low-pressure gas to generate electron-ion pairs. An electric field then transports the electrons to a readout plane where they are amplified and collected by the MPGD. \mutpc{}s can therefore record a recoil track in three dimension. While three-dimensional tracking is possible with this approach, diffusion limits the drift length of the detector. To suppress diffusion and improve track reconstruction, a negative-ion drift gas can be used, in which ionization electrons attach to electronegative gas molecules to form negative ions \cite{martoff}. The negative ions are then transported to the readout plane. These TPCs are called negative-ion (NI) \mutpc{}s.  The {\it DRIFT} experiment uses a NI TPC, and has demonstrated diffusion at the thermal limit \cite{DRIFT2}.

In addition to reduced diffusion, the use of a NI \mutpc{} also enables full-volume fiducialization through 3D position reconstruction of the event vertex \cite{minority}.
Because the main backgrounds of TPC-based dark matter searches are $\alpha$-particles and daughter nuclei from $^{238}$U/$^{232}$Th chain decays of isotopes on the MPGD and drift planes, fiducialization can enable background-free operation \cite{DRIFT2, DRIFT1}. Fiducialization along the drift direction is possible because NI gases can produce multiple species of charge carriers, each with different mass and therefore mobility. As a result, the difference of arrival times between species correlates with total drift distance. For example, nuclear recoils in SF$_6$ primarily produce SF$_6^-$, but also trace amounts of SF$_5^-$ (so-called a ``minority carrier''), whose mobility is $\approx 5\%$ larger \cite{phan}. Because the time-scale for events in a NI \mutpc{} is orders of magnitude longer than for an electron-drift TPC, new electronics, such as the \textit{LTARS}, are required to read out these detectors.


In addition to improving the sensitivity of directional dark matter searches, the \textit{LTARS} system can also benefit large-scale liquid argon (LAr) detectors. An $\mathcal{O}$(10~kt) LAr time projection chamber (LAr-TPC) will be utilized for next-generation neutrino oscillation studies, and also for studies of nucleon decay and astrophysical neutrinos.
Toward the realization of such a large detector,
a world-wide R\&D effort on kilo-ton-scale LAr-TPC demonstrators is underway \cite{WA105}.
One approach to reading out the ionization electron signal uses a dual-phase LAr-TPC \cite{LEM}.
In the dual-phase TPC, the ionization electrons are extracted from the liquid to the vapor phase, and extracted electrons are multiplied with a thick-GEM (Gas Electron Multiplier) and collected with a two-dimensional strip anode. 
The main advantage of a dual-phase readout is the high signal-to-noise ratio afforded 
by the gas multiplication. This enables longer drift lengths over which some primary electrons are lost to impurities in the LAr.
The high signal-to-noise ratio also benefits the 
physics sensitivity for the 
neutrino oscillation, nucleon decay, and astrophysical neutrino signals.
%

Both the directional dark matter search and LAr TPCs for neutrino experiments require innovations in readout electronics.
Since the drift speed of negative-ions is 
orders of magnitude smaller than that of electrons in a low-pressure gas ($\approx 10^{-2}$ cm/$\mu$s \vs{} $\approx 1$ cm/$\mu$s), existing electronics designed for high-rate TPC readout in collider experiments are too fast to collect charges from NI \mutpc{}.  Additionally, since the minority carrier signal strength is typically only a few percent of the primary signal
in optimized gas conditions, a wide dynamic range is also required for the electronics to sense SF$_6^-$/SF$_5^-$ with high signal-to-noise ratio (\eg{}, typical signals are 
3 fC for minority species and 80 fC for the main species).
Also, fine MPGD spatial resolution necessitates a high-channel count readout in a small form-factor (\eg{}, 1500 channels for a $30\times30$ cm$^2$ readout area with strip pitch of 400 $\mu$m).
The signal timescale for LAr-TPCs is similar to that of the NI \mutpc{} since the ionization electrons drift
in the liquid (the drift speed is typically $\approx 10^{-1}$ cm/$\mu$s).
The wide dynamic range requirement also applies to the LAr-TPC. 
The typical ionization signal for a minimum ionization particle (MIP) track 
is $\approx 10$ fC depending on the thick-GEM gain.
A single-channel signal for
events associated with electromagnetic or hadronic showers
is $\sim$50 times larger than for a MIP.
The number of readout channels is expected to be $\mathcal{O}(10^5)$ for 
a $\mathcal{O}(10\,{\rm kt})$ detector.
Moreover, to minimize detector capacitance, the analog front-end electronics
must be located as close as possible to the strip readout.
As a consequence, the analog front-end must operate at $\sim$88 K.

In this article, we report on the recent development and characterization of a prototype analog front-end ASIC, which meets the needs of both NI \mutpc{} and LAr-TPC detectors.
 We describe the ASIC design in section 2, report on the test results in section 3, discuss the design refinements in section 4, and give conclusions in section~\ref{sec:discussion}. 
 The characterization of 88~K operation is out of scope of the present paper. The full detail of the cryogenic operation is going to be described in a next paper.

%
%
\section{Specification of the ASIC}
\subsection{Overview of the ASIC}
\begin{figure}[!t]
\centering
\includegraphics[width=2.4in, angle=180]{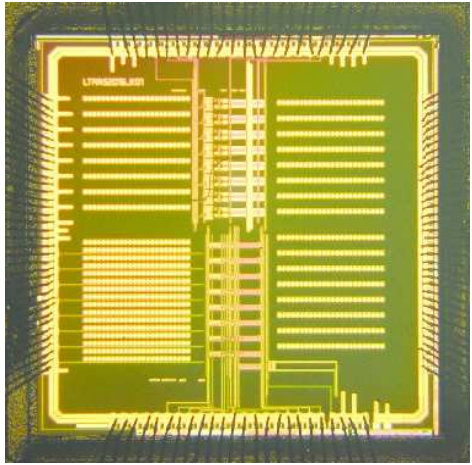}
\caption{\label{fig_foto}Photograph of the readout ASIC. The chip size is $5\times5$~mm$^2$. A total of 16 readout channels are located in the central region, and the remainder of the chip houses decoupling capacitors to stabilize the bias and power supplies.}
\end{figure}

The readout ASIC was implemented with Silterra 180-nm CMOS technology with a chip size of 5 $\times$ 5 mm$^2$. The objectives of the prototype were to establish operation of basic signal processing components such as a preamplifier and CR-RC shaper stage, and to demonstrate the successful operation of these elements on a detector.
For evaluation purposes, we implemented two different signal processing chains in the ASIC: (a) a pair of static gain stages to handle wide ($\pm$1600~fC) and narrow ($\pm$80~fC) dynamic-range signals (hereafter referred as ``static architecture''), and (b) a dynamic gain-switching architecture (hereafter ``dynamic architecture''). Each ASIC implements eight independent channels of each architecture.
Table \ref{tab1} lists the specification and performance requirements of the ASIC, and Figure~\ref{fig_foto} shows a photograph of the ASIC. Signals are processed from left to right on the chip. The main circuit core occupies the central quarter of the ASIC. The remainder of the chip houses bias nodes for the two architectures and decoupling capacitors.

\begin{table*}
\begin{center}
\caption{\label{tab1} Specification and requirements of the ASIC.}
\small
  \begin{tabular}{|c|c|c|} \hline
Technology & \multicolumn{2}{|c|}{Silterra 180~nm CMOS} \\ \hline 
Chip size & \multicolumn{2}{|c|}{5$\times$5~mm$^2$, 16 total channels} \\ \hline
Supply power & \multicolumn{2}{|c|}{1.8~V core/IO, $\pm0.9$~V operation, max. 2.4~mW/ch}\\ \hline
Fabrication options &\multicolumn{2}{|c|}{6 metals, deep N-well, high-value poly res., MIM cap.} \\ \hline \hline
Minimum signal charge & 3 fC (minority species) & 100 fC (main species)\\ \hline
ENC & 2000~e$^-$ (S/N=10) & $< 6.4 \times 10^4$ e$^-$ \\ 
        & 4000~e$^-$ (S/N=5, see Section~\ref{sec:discussion}) & \\ \hline
 Dynamic range & $\pm$80 fC (narrow range)& $\pm$1600 fC (wide range)\\ \hline
Voltage gain & 10~mV/fC & 0.5~mV/fC \\  \hline
Shaping time & \multicolumn{2}{|c|}{4--7~$\mu$s for NI $\mu$TPC / 1--4~$\mu$s for LAr-TPC} \\ \hline
  \end{tabular}
\end{center}
\end{table*}

\subsection{Circuit Properties}
\subsubsection{Static gain architecture}
\begin{figure*}[!t]
\centering
\includegraphics[width=5in]{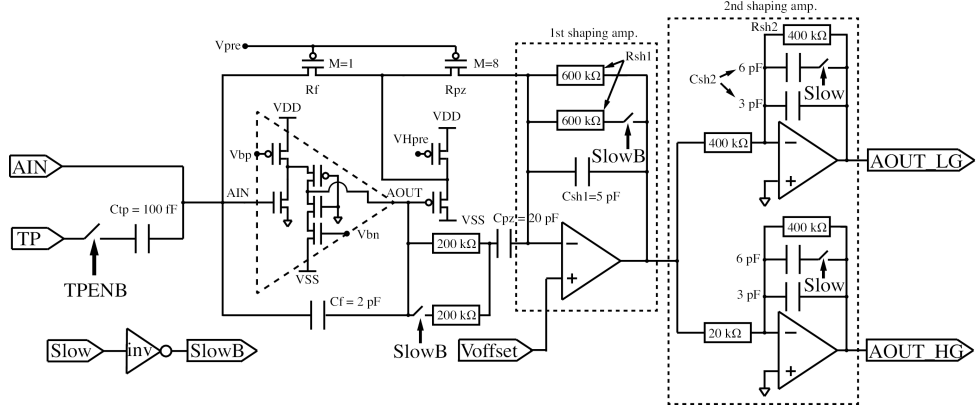}
\caption{Block diagram of the static gain architecture. $R_{\rm f}$ and $R_{\rm pz}$ are within the range of several M$\Omega$ adjusted by the gate voltage $V_{\rm pre}$.}
\label{fig_MT}
\end{figure*}

Figure~\ref{fig_MT} shows the block diagram of a single channel of the static architecture. 
The detector is externally AC-coupled with electro-static discharge protection diodes to the charge-sensitive amplifier (CSA) input.  Calibration test pulses can be injected via an on-chip  capacitor of 100~fF. 
The CSA implements a modified folded cascode topology with a large input NMOS transistor (W/L = 40~$\mu$m/300~nm, M = 100, g$_{\rm m}$/I$_{\rm d}$=7.5~mS/280~$\mu$A). The bias condition of input and load transistors are adjusted independently by acquiring constant currents from the VDD and VSS.
A transfer gate-type PMOS is employed for the CSA DC-feedback component. 
The CSA is designed to handle signals of either polarity, and includes pole-zero cancellation.
The CSA drives a 2-stage shaping amplifier, whose output is shared by high- and low-gain amplifiers whose gains are optimized for narrow and wide dynamic-range signals (see Table \ref{tab1}). Although the time constants of each stage in Fig. \ref{fig_MT} seem to be set at 3~$\mu$s/1.5~$\mu$s ($C_{\rm sh1}R_{\rm sh1}$) and 3.6~$\mu$s/1.2~$\mu$s ($C_{\rm sh2}R_{\rm sh2}$), respectively, the overall response becomes slower since the PZC (pole-zero cancellation circuit) is directly connected to the CSA output instead of the source follower's output as a popular configuration. We thus carefully designed for a default peaking time of 7~$\mu$s/$3.5~\mu$s by SPICE simulation. In this work, we focus on the 7~$\mu$s performance.
All resistor components are implemented with high-value poly resistors (1.05~K$\Omega$/sq.).

\subsubsection{Dynamic gain-switching architecture} 
\begin{figure*}[!t]
\centering
\includegraphics[width=5in]{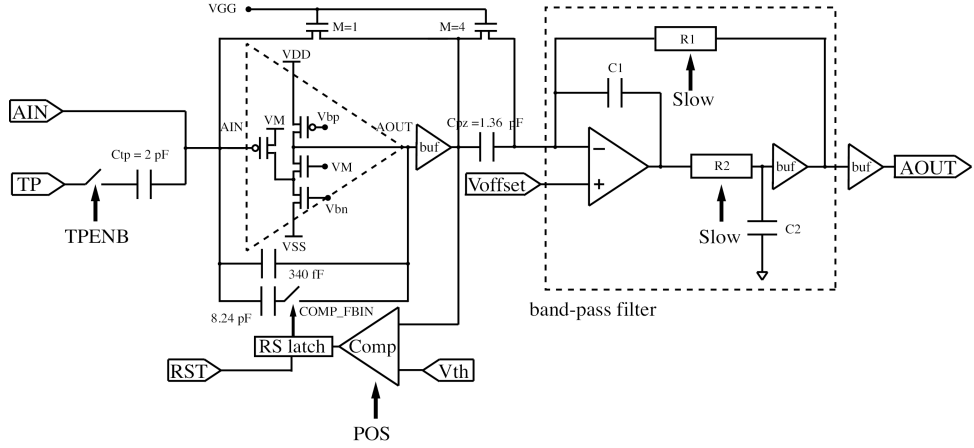}
\caption{Signal processing chain of the dynamic gain switching architecture. Typical values of resistors and capacitors in the band-pass filter are $R_1=R_2=46.3$/9.3~M$\Omega$, $C_1=216$~fF (=$4\times C_2$), and $C_2=54$~fF.}
\label{fig_TK}
\end{figure*}

The signal processing chain for the dynamic gain switching architecture is shown in Figure~\ref{fig_TK}. The circuit is based on Ref.~\cite{geronimo}.
The CSA uses a folded cascode configuration with an input PMOS transistor, and the CSA DC-feedback FET is implemented with NMOS.
The feedback capacitor is selectable via a CMOS switch (designated as \texttt{COMP\_FBIN}) between 340 fC and 8.58 pF for narrow-range and wide-range modes, respectively. This gain setting of either ``\texttt{L}'' or ``\texttt{H}'' (at first set as ``\texttt{H}'' with high-gain) is provided from a discriminator output after comparison with the CSA output and an external threshold voltage. An \texttt{RS latch} is inserted between the gain switch and the discriminator to prevent unexpected bit transition during the signal processing. After finishing each event process, the latched gain setting is reset from an external DAQ (or FPGA) board. The CSA output is fed into a band-pass filter circuit with pole-zero cancellation and a second-order low-pass filter. The latter utilizes two sets of high-resistance circuits, functioning as a selectable mega $\Omega$ resistor by implementing with a poly resistor of 66/330~k$\Omega$ and current mirrors \cite{kishishita}.
The transfer function of the band-pass filter (indicated in the dashed line in Figure~\ref{fig_TK}) is:
\[ T(s)=-\frac{R_1}{(s^2C_1C_2R_1R_2+sR_1C_1+1)}.
\]
By choosing  $\displaystyle C_1R_1=4C_2R_2$, the transfer function becomes $\displaystyle T(s)=-\frac{R_1}{(2sC_2R_2+1)^2}$, resulting in the second order low-pass filter. The peaking times are selectable via the CMOS switch of the resistor circuits $R_1$ \& $R_2$.
Circuit configuration, such as test pulse enable and peaking time selection, is done by daisy chained shift registers.

\section{Measurement Results}
\subsection{Experimental setup}
To evaluate the circuit properties, the ASIC was directly mounted on a test board before connecting to a detector. A computer interface was established using a commercially available digital board with a Xilinx FPGA (the SOY board from Bee Beans Technology \cite{bee}). All necessary bias voltages for the ASIC were provided from potentiometers and resistors on the PCB.

\subsection{Waveforms}
\begin{figure*}[!t]
  \centering
    \begin{tabular}{c}

      \begin{minipage}{0.48\hsize}
          \includegraphics[keepaspectratio, scale=0.4, angle=0]
                          {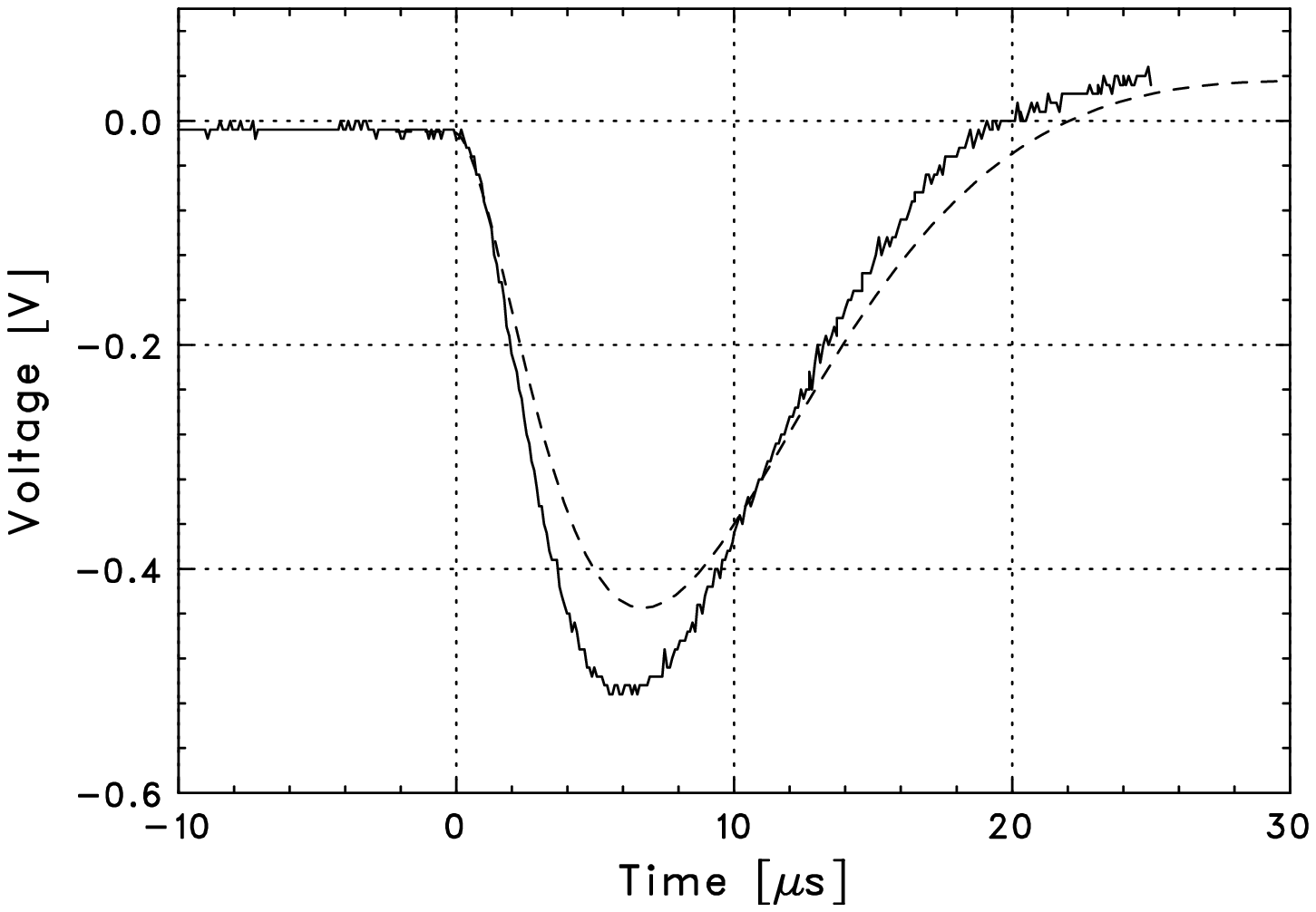}
       \end{minipage}
 
 
      \begin{minipage}{0.02\hsize}
        \hspace{2mm}
      \end{minipage}

      \begin{minipage}{0.48\hsize}
          \includegraphics[keepaspectratio, scale=0.4, angle=0]
                          {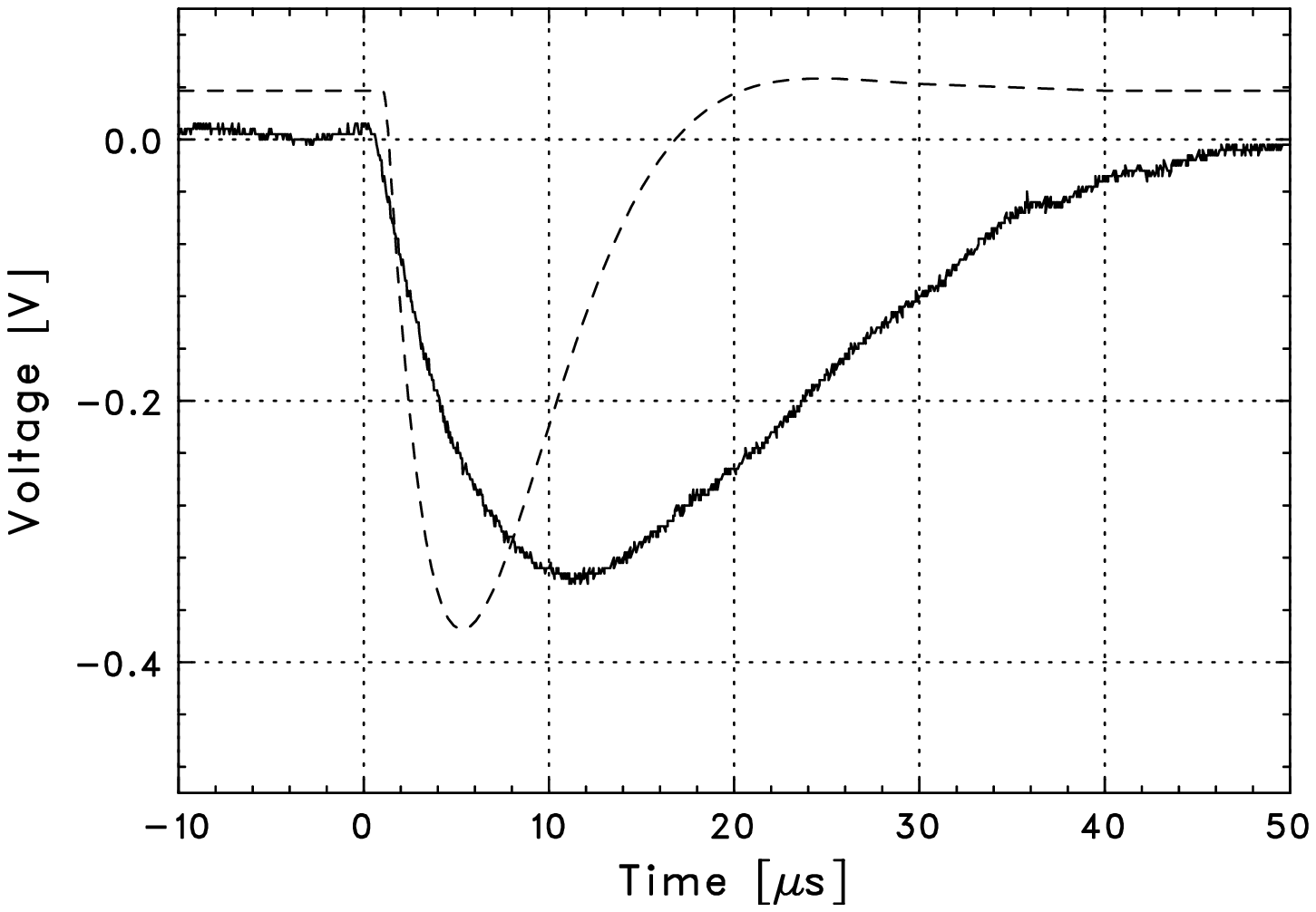}
      \end{minipage} \\
    \end{tabular}
                             \caption{Output comparison of the static (left) and dynamic (right) architectures in the slow peaking time mode: dashed lines indicate SPICE simulation and solid lines show measurement results. Injected charges are both 50~fC with a pseudo-detector capacitance of 300~pF. The gain setting was fixed to high-gain mode for the dynamic architecture. }
                          \label{fig_waveform}

\end{figure*}

Figure~\ref{fig_waveform} shows the high-gain outputs for each architecture for an input charge of $-50$~fC. In the static architecture (left panel), the peak pulse height is measured to be $-512$~mV for a peaking time of $5.4~\mu$s. The simulation predicts a smaller pulse amplitude ($-426$~mV) and a slightly slower peaking time ($6.7~\mu$s).

As for the dynamic architecture, we needed bias optimization before injecting test pulses. This is because the default bias settings obtained from simulation did not provide appropriate conditions to the ASIC, and we observed the following phenomena: (1) the measured pulse height was 2--3 times smaller than the simulation for the default bias voltages, and (2) the effect of the detector capacitance $C_{\rm det}$ was worse than predicted, \ie{} the gain decreases rapidly as $C_{\rm det}$ increases. We describe these issues in detail in Section~\ref{sec:discussion}. After increasing the bias voltage on the \texttt{VM} node by 0.5~V (see Figure~\ref{fig_TK}), analog outputs were observed from the chip, as shown in the right panel of Figure~\ref{fig_waveform}. Even after bias optimization, however, the measured waveform peaking time ($\sim11$\,$\mu$s) does not match expectation, and the pulse height was still small by 10\%.  

\begin{figure*}
\centering
\includegraphics[width=5in]{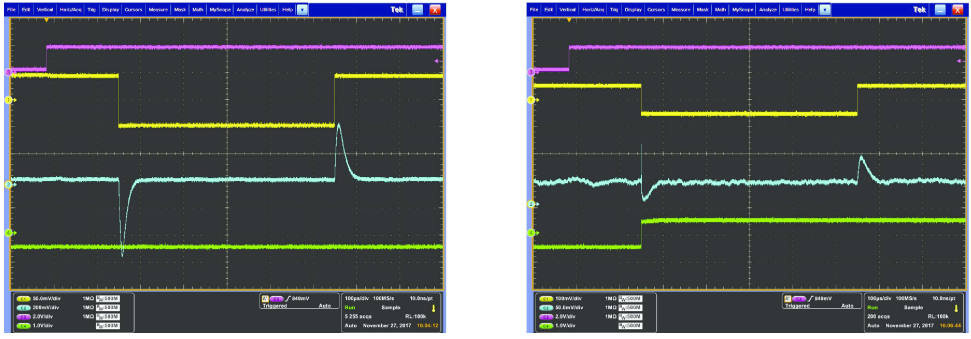}
\caption{Waveforms of the dynamic gain-switching response for input charges of 90~fC (left) and 100~fC (right). The signals from top to bottom are reset signal (magenta), test pulse from a function generator (yellow), shaper output (cyan), and \texttt{COMP\_FBIN} (green). }
\label{fig_oscillo_dyn}
\end{figure*}

Next, we characterized the dynamic gain-switching response. Figure~\ref{fig_oscillo_dyn} shows screen shots for injected charges of 90~fC and 100~fC, respectively. The threshold level was set between those values. We can clearly see that the output of \texttt{COMP\_FBIN} (indicated in green at the bottom) in the left figure remains in the low state below the threshold, while the expected transition is seen in the right panel when the shaper output crosses threshold. A short spike in the shaper output of the right panel seems to coincide with a rising edge of \texttt{COMP\_FBIN}, however, this is an artifact caused by noise injection due to coupling between oscilloscope probes. The switching noise from the CMOS switch seems clearly removed in the band-pass filter, and the waveform shows the expected semi-gaussian shape.


%


\subsection{Voltage gain and Dynamic range}  
\begin{figure*}[!t]
  \centering
    \begin{tabular}{c}

      \begin{minipage}{0.48\hsize}
          \includegraphics[keepaspectratio, scale=0.5, angle=0]
                          {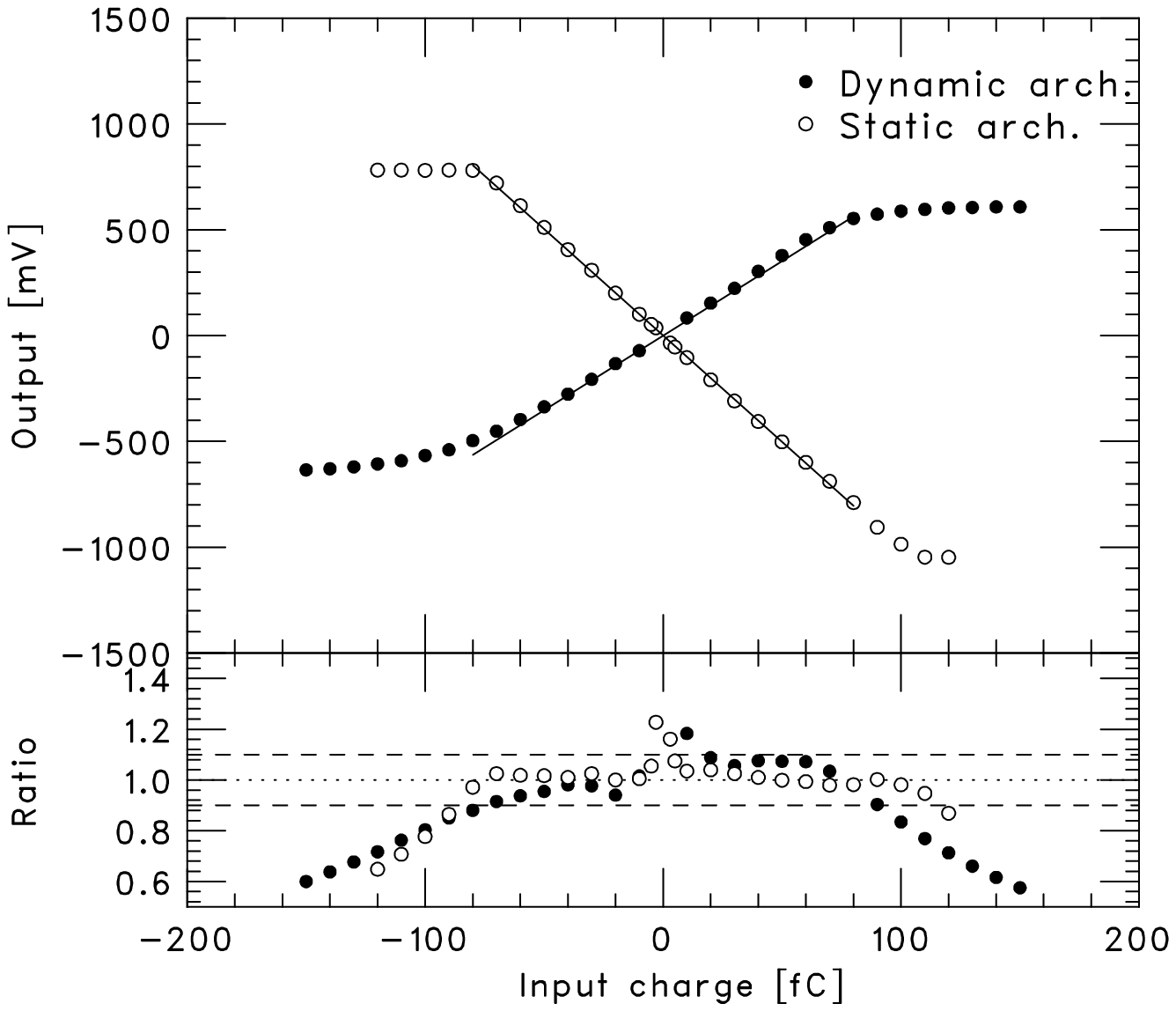}
       \end{minipage}
 
 
      \begin{minipage}{0.02\hsize}
        \hspace{2mm}
      \end{minipage}

      \begin{minipage}{0.48\hsize}
          \includegraphics[keepaspectratio, scale=0.5, angle=0]
                          {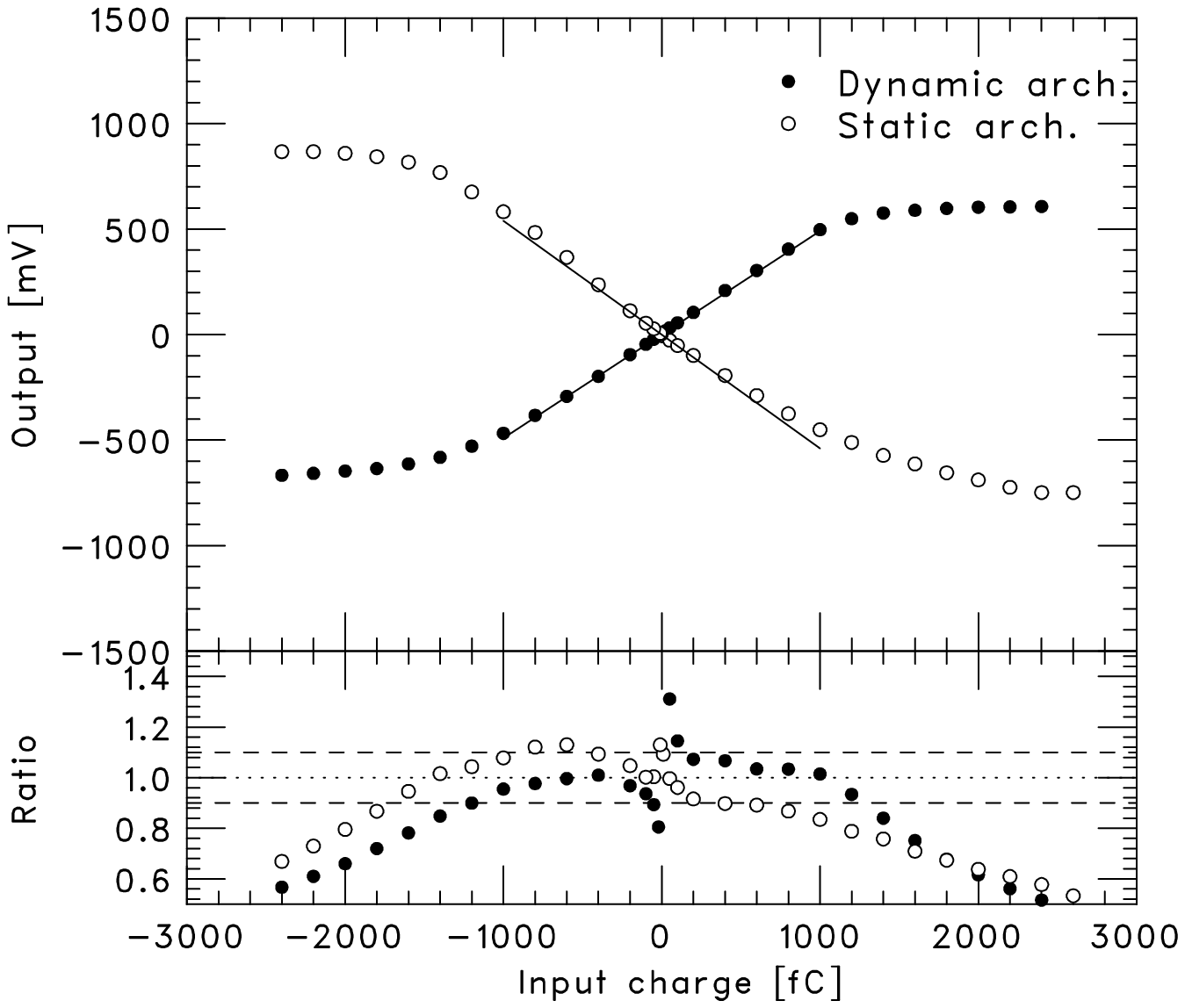}
      \end{minipage} \\
    \end{tabular}
                             \caption{Dynamic range and linearity curves of the high-gain (left) and low-gain (right) modes. Black and white points indicate dynamic and static architectures, respectively.}
                          \label{fig_lin}

\end{figure*}

Figure~\ref{fig_lin} shows measurements of the dynamic range for each architecture. The lower panels show the ratio of the measurements and a linear fit between $\pm 80$~fC for high-gain and $\pm 1000$~fC for low-gain, respectively. A $\pm 10$\% level is also presented in dashed lines. This level is defined as a negligible range, compared with an energy resolution of the gas detectors of typically $20$\%, including a position dependence of the MPGD \cite{nakamura2}.
For the static architecture, the measured gains are 10.05~mV/fC and 0.54~mV/fC, respectively, which meet the \textit{LTARS} design specification. For the fast peaking time mode (3.5\,$\mu$s), the measured gains are 10.29~mV/fC and 0.51~mV/fC, which also meet the design specification. For the dynamic architecture, the measured gains are 7.05~mV/fC and 0.49~mV/fC (7\,$\mu$s peaking time), and 7.33~mV/fC and 0.51~mV/fC (1\,$\mu$s peaking time).

The low-gain dynamic ranges for both architectures look narrower than the design specification of $\pm 1600$~fC. However, by adjusting the output baseline offset, the specification can be achieved. For standard operation, this is a satisfactory approach because the ASICs operate with a pre-determined polarity.

\subsection{Noise characteristics}
\begin{figure}[!t]
\centering
\includegraphics[width=3in]{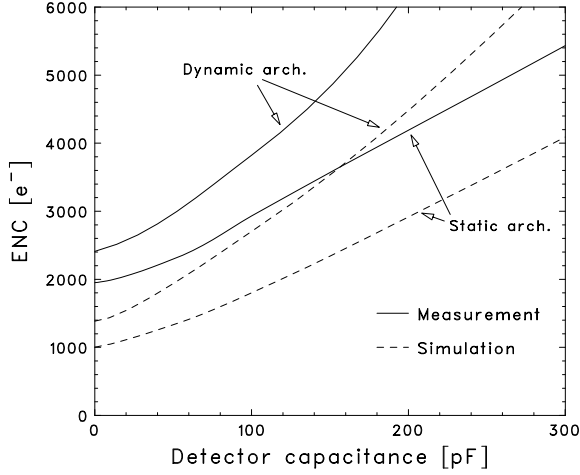}
\caption{ENC in the high-gain mode as a function of detector capacitance.}
\label{fig_noise}
\end{figure}

The simulated and measured equivalent noise (rms) as a function of detector capacitance is shown in Figure~\ref{fig_noise}. The simulation includes parasitic resistance and capacitance of the core circuits. The noise slopes are roughly consistent with the simulation for both architectures, however, we can clearly see a noise offset of $\sim$1000~e$^-$. 
The origin of that excess noise offset is under investigation, however the 2\,mm-long metal traces with a width of 3~$\mu$m between the ASIC pads and the CSA inputs are likely contributing and/or easily affected by external noises. On the other hand, the measured ENCs of the low-gain modes meet the design requirements for a detector capacitance of 300\,pF: 6950~e$^-$ for the static architecture and 3.5$\times10^4$~e$^-$ for the dynamic architecture. Although the noise measurement for the dynamic architecture was made without activating the gain-switching circuit, the noise contribution from CMOS switching is negligibly small ($\sim$3\%) compared with the ENC in the low-gain mode.

%
%
%
%
%

\section{Demonstration with a Micromegas detector}
\subsection{Experimental setup}
\begin{figure*}[!t]
  \centering
    \begin{tabular}{c}

      \begin{minipage}{0.4\hsize}
          \includegraphics[keepaspectratio, scale=0.4, angle=0]
                          {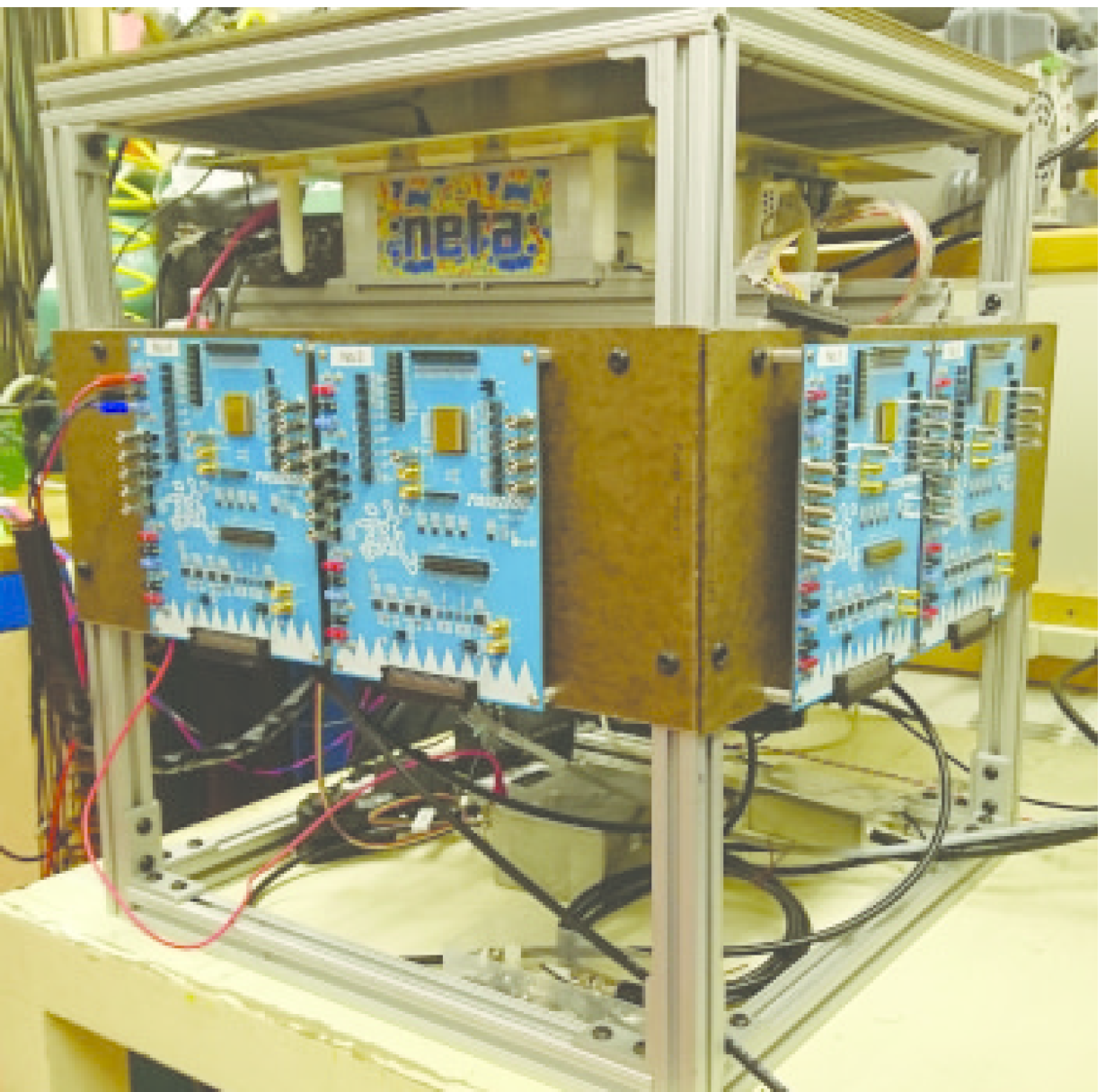}
       \end{minipage}
 
%
%
 
      \begin{minipage}{0.43\hsize}
          \includegraphics[keepaspectratio, scale=0.3, angle=0]
                          {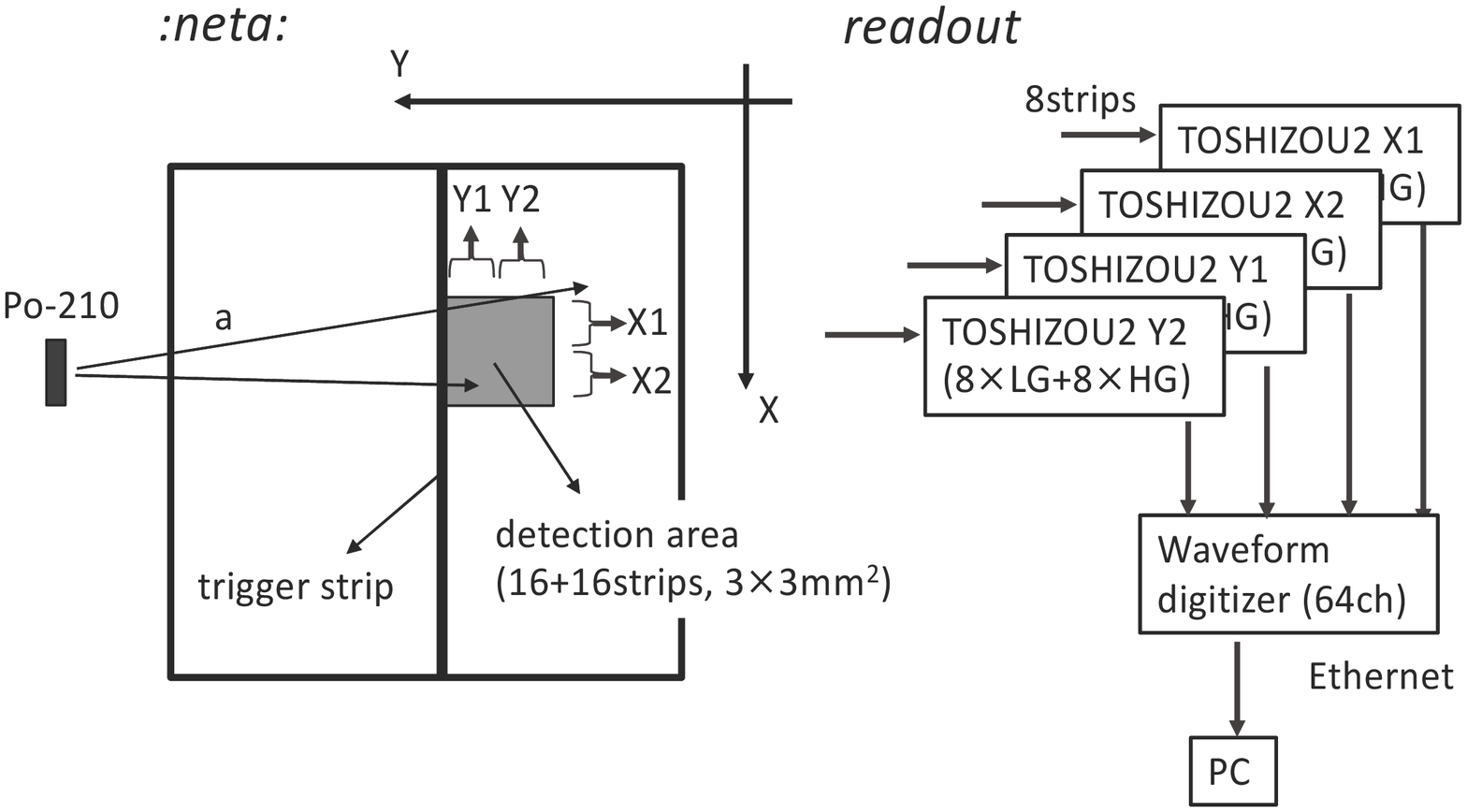}
      \end{minipage} \\
    \end{tabular}
                             \caption{Experimental setup (left) and schematic readout diagram (right). Two of the four TOSHIZOU2 test boards (blue) mounted on the lateral sides of the setup are visible in the photograph.}
                          \label{fig_setup1}

\end{figure*}

After characterizing the full chip functionality, we connected the \textit{LTARS} test boards to a NI \mutpc{} with Micromegas readout to assess the performance on a detector.
The detector was developed as a prototype directional dark matter detector at Wellesley College in the USA. 
The experimental setup and the schematic readout diagram are shown in Figure~\ref{fig_setup1}. 
The experiment was performed with an $\alpha$-ray source and the tracking capability for $\alpha$-rays was demonstrated.

The Micromegas was made by CERN with an amplification area and gap of 10 $\times$ 10 $\rm cm^2$ and 256 $\mu$m, respectively \cite{alex}. 
The drift length was 1.5 cm and a radioactive source of $^{210}$Po was placed 5~mm away from the Micromegas.
SF$_6$ gas with a pressure of 60~Torr was used as a NI-drift gas. The high voltages of the drift plane and mesh were $-1525$\,V and $-625$\,V, which formed a drift electric field of 0.6~kV/cm. The gas gain was $\sim$100. A two-dimensional strip readout with a high-resistivity strip layer for spark protection on the top was used. The pitch of the readout and protection strips was 200~$\mu$m. The strip patterns were formed on a film with a thickness of 50~$\mu$m. Three films, namely Y-strips (bottom), X-strips (middle) and resistive-strips (top), were layered. The resistive strips were parallel to the Y-strips. The $\alpha$-source was mounted such that the mean $\alpha$-particle direction was parallel to the X-strips. A total of 32 strips (16 consecutive X-strips and 16 consecutive Y-strips) covering a detection area of $3.2\times3.2$~mm$^2$ were connected to four test boards, labeled as TOSHIZOU2 in Figure~\ref{fig_setup1}. The closest Y-strip to the $\alpha$-source was used as a trigger. The analog outputs from the test boards were digitized at 2.5~MHz by the FPGA board. The resulting digital data was recorded by a PC. All data were taken with the static gain architecture in the high-gain, slow peaking time mode to reduce electronic noise. The noise levels were 16~mV$_{\rm rms}$, corresponding to an ENC of 9937~e$^{-}$. The detector capacitance per strip (10 cm long) was measured to be 50 pF. This value is approximately four times larger than the estimation from Fig. 7. However, periodic noise is clearly visible in the waveforms of Fig. 10, indicating that the remaining noise originates elsewhere in the experimental setup.

 \subsection{Track reconstruction}
\begin{figure}[!t]
\centering
\includegraphics[width=3.4in]{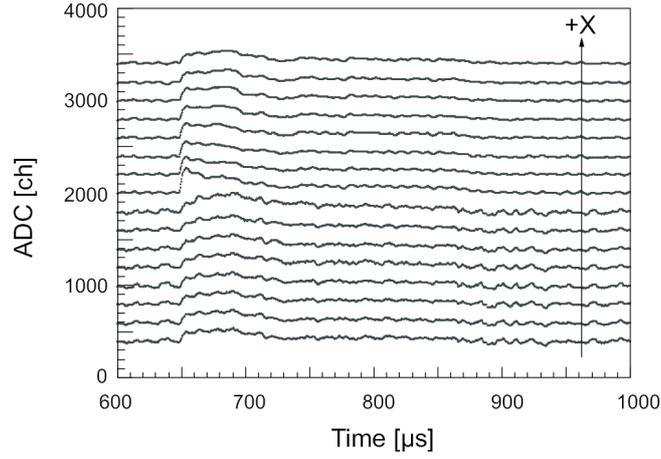}
\caption{Waveforms from the 16 instrumented X-strips. The baseline for each channel is offset for clarity.}
\label{fig_timingX}
\end{figure}

\begin{figure}[!t]
\centering
\includegraphics[width=3.4in]{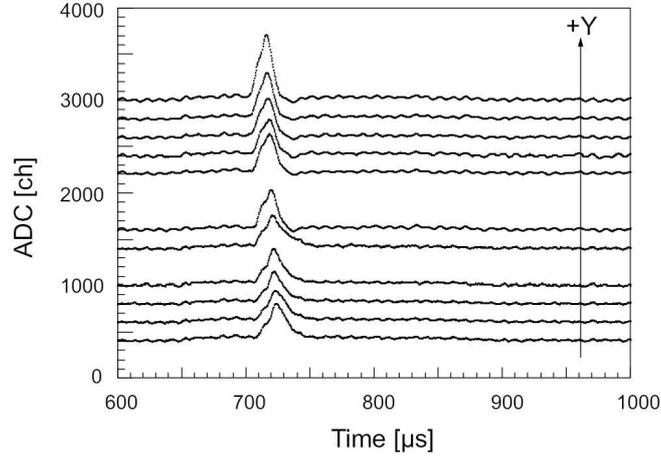}
\caption{Waveforms from 11 instrumented Y-strips (5 channels were not functioning). The baseline for each channel is offset for clarity. The upper strip is closest to the $\alpha$-source, and peaks earliest in time. This track thus pointed away from the Micromegas.}
\label{fig_timing}
\end{figure}

 Figures~\ref{fig_timingX} and \ref{fig_timing} show sample waveforms from each strip for a single $\alpha$-event. Some channels in the Y-strips were broken during the measurement. In this two-dimensional readout with resistive strips on the top, positive charges are induced when the amplified electrons are collected at the high-resistive strip. Thus, the polarities of observed waveforms from both strips are consistent with expectation. While we can see the TPC-like signal propagation of fast signals
clearly in the Y-strips (top waveforms first and bottom waveforms last), the
induced signals in X-strips are much longer in duration than Y-strips. The X-strip waveforms
seen in Figure 9 have long duration because charges propagate in the high-resistive strips
(parallel to the Y-strips) with a time constant of $\mathcal{O}$(10$\sim$100) $\mu$s and induce extra hits in the X-strips.
Although three-dimensional track reconstruction was hindered due to this effect, we successfully demonstrated that the \textit{LTARS} ASIC could successfully read out signal charges from the Micromegas with a very low gain of $\sim$100 and a strip pitch of 200~$\mu$m. 
 
\begin{figure}[!t]
\centering
\includegraphics[width=3in]{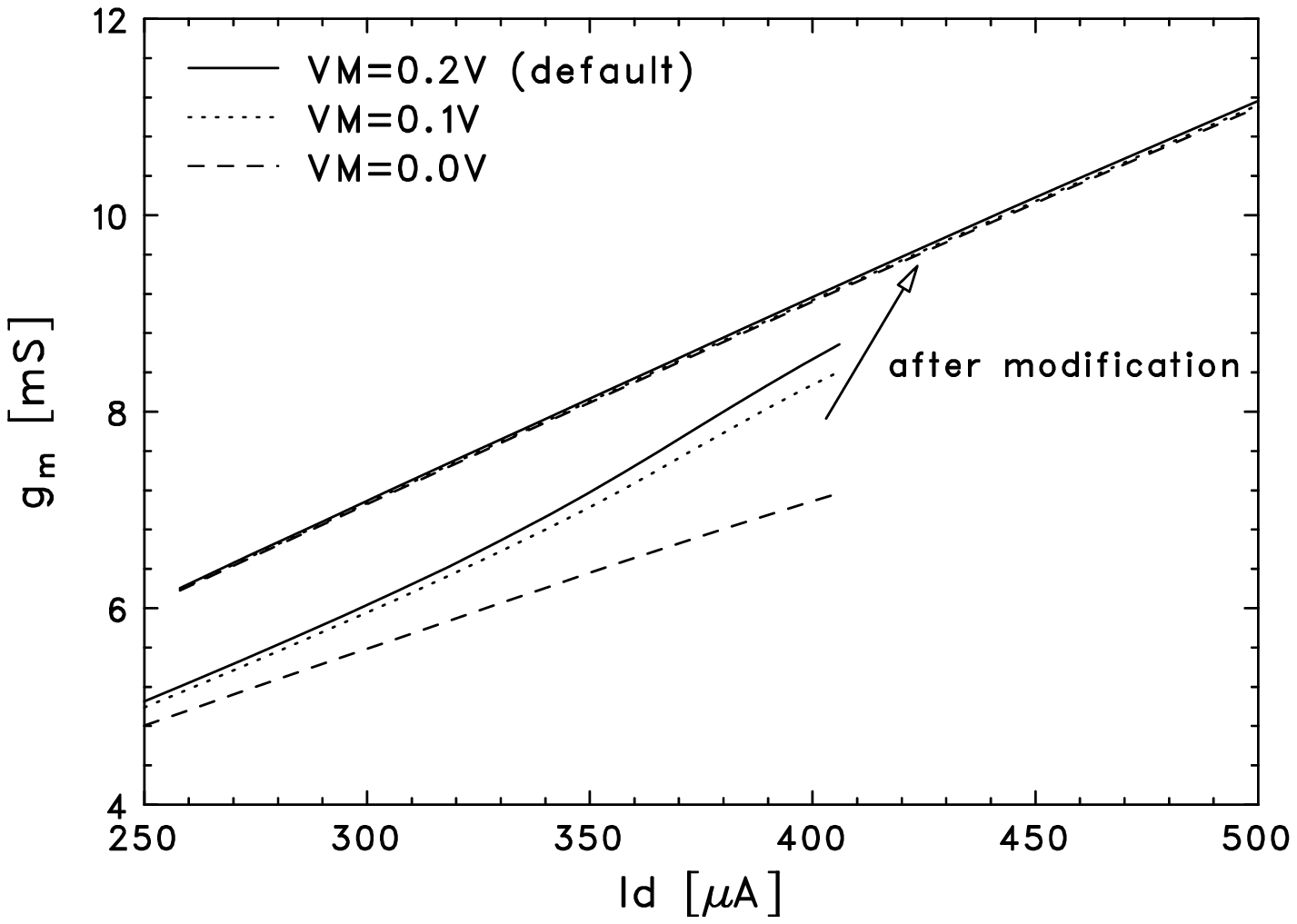}
\caption{$g_{\rm m}$ \vs{} $I_{\rm d}$ of the input PMOS in the dynamic architecture.}
\label{fig11}
\end{figure}

\section{Discussion for performance improvements}
\label{sec:discussion}
In this section, we address the discrepancy between the measured and simulated performances for each architecture, and discuss design improvements for the next version of the \textit{LTARS} ASIC. The static architecture meets the design specification, but with slightly larger gain than expected (see Figure~\ref{fig_waveform}, left). This discrepancy cannot be explained by the transistor's corner parameters. We speculate instead that process variation of the poly resistor is the cause, \ie{}, three high-value poly resistors are placed in series with the minimum design width and dummy structures for a large resistance of 600~k$\Omega$ in the first shaping amplifier ($R_{\rm sh1}$ in Fig. \ref{fig_MT}). Since the poly resistors are formed by chemical etching on metal layers in this technology, the absolute resistance accuracy is typically 10\% even with dummy structures, when the poly width follows the minimum rule.
If we denote $R_{\rm sh1}$ as the feedback resistor in the first shaping amplifier (see Fig. \ref{fig_MT}), the voltage gain which is proportional to $R_{\rm sh1}$ becomes larger than expected under over-etching condition.

On the other hand, the performance discrepancies of the dynamic architecture cannot be explained only by the process variation since the circuit does not include poly resistors over 330~k$\Omega$ in the high-resistance circuit. The origin is likely due to the transconductance $g_{\rm m}$ of the input PMOS in the preamplifier ($W/L=11.2~\mu$m/ $L=200$~nm, $M=100$). The hypothesis that the dynamic architecture performance is affected by $g_m$ also explains other observations. Due to the Miller Effect, the effective input capacitance of the amplifier is $C_f (1+A)$, where $C_f$ is the feedback capacitance and $A$ is the open-loop gain of the amplifier. Hence, the fraction of charge generated in the detector and absorbed by $C_{\rm f}$ is written as $\displaystyle \frac{C_{\rm f}\cdot(1+A)}{C_{\rm det} + C_{\rm f}\cdot(1+A)}$, where $A=g_{\rm m}\times r_{\rm o}$, and $r_{\rm o}$ is the output impedance of the preamplifier. Ideally, $g_{\rm m}$ is sufficiently large, and the $C_{\rm det}$ dependence can be ignored. The $g_{\rm m}$ that is smaller than intended will degrade the S/N.
This is consistent with the observation in the $C_{\rm det}$ \vs{} gain curve, which was worse than the design. A small $g_{\rm m}$ also affects the output peaking time. The rise time of the CSA is generally given as $t_{\rm r}=\frac{C_{\rm det}}{g_{\rm m}}+\frac{C_{\rm L}}{\mu_0 g_{\rm m}}$, where $\mu_0\equiv\frac{C_{\rm f}}{C_{\rm f}+C_{\rm det}}$, and $C_{\rm L}$ is a capacitive load at the output node of the preamplifier. With sufficiently large $g_{\rm m}$ (ideal case), the rise time of the preamplifier is fast, while the rise time becomes longer with decreasing $g_{\rm m}$ (current case). As a result, a slower waveform is observed at the final output stage (Figure~\ref{fig_waveform}, right). This is caused through multiple factors: (i) the expected $I_{\rm d}$ \ie{} designed as $I_{\rm}$=500~$\mu$A, is not supplied to the PMOS even under the maximum external bias, due to insufficient transistors' $W/L$ in bias current mirrors, and (ii) $g_{\rm m}$ is sensitive to the VM node since the node is not only connected to the source of the input FET, but also to the gate of the cascode transistor (Fig. \ref{fig_TK}). If we consider the IR drop in the VM node, the decrease of $g_{\rm m}$ is a reasonable conclusion. Figure \ref{fig11} shows the simulation result of $g_{\rm m}$ \vs{} $I_{\rm d}$ of the input PMOS under various VM condition. After the optimization of the bias current mirrors, we separated the gate of cascode FET from the VM line, and also increased the finger number of the cascode FET. The improved $g_{\rm m}$ is overlaid in Fig. \ref{fig11}.

The remaining issue is the ENC. This should be considered from experimental viewpoints. As shown in Table~\ref{tab1} and Figure~\ref{fig_noise}, the targeted noise level of less than 2000~$e^{-}$ (rms) was not achieved in the current design. 
If we relax the $S/N$ constraint to 5, fake events appear at 0.2 Hz in one strip, taking account of the band-width of $\sim$ 100~kHz. In the \mutpc{} experiment, we normally read
out multiple strips for each event, and the fake events can be easily recognized
if the noise rate is about 0.2 Hz.  We thus reset the design specification of ENC to 4000~e$^-$ ($S/N$=5) and we will improve the layout to achieve the design specification in the next submission.


\section{Conclusion}
For the purpose of reading out NI \mutpc{} and LAr-TPCs, we have newly developed a dedicated analog ASIC in 180-nm CMOS technology. To meet the wide dynamic range of 1600~fC with a detector capacitance of 300~pF, the ASIC contains two different readout architectures: one was constructed with a static and separate gain-stages and another was implemented with a dynamic gain-switching topology. We mounted the ASIC on a test board and compared the ASIC performance to a SPICE simulation, in terms of shaping time, dynamic range, and ENC.
For the static architecture, the shaping time and dynamic range meet the design specifications. For the dynamic architecture, the gain switching works as designed, however, some design issues were found in the preamplifier circuit. By using a multi-finger cascode FET and separating the source of the input FET and gate of the cascode FET in the preamplifier, we expect to fix the problem, and achieve the required noise level in the next version of the ASIC. The ENCs of both architectures are still larger than desired, however, the excess noise is likely due to the long metal routing between a wirebonding pad and an analog input. This issue can be improved by optimizing the layout. In anticipation of cryogenic operation, we carefully chose the gate lengths of devices to reduce the impact ionization probability at 88 K. At low temperature, the CMOS transconductance increases due to the increased carrier mobility, which is expected to improve the noise performance. In order to assess such cryogenic effects, however, the overall noise performance must be improved.  Therefore the cryogenic characterization of the LAr-TPC will be done with the next design.
We conclude that the static architecture reaches the requirements for an actual NI \mutpc{} dark-matter experiment. We also demonstrated the ASIC front-end on an NI \mutpc{} with Micromegas gas amplification and 200 $\mu$m-pitch strip electrodes.

\acknowledgments
This work was supported by the JSPS KAKENHI (Grant No. 25105008, 16H02189, 26104005, 15K21747, 17H04840), and the U.S.-Japan Science and Technology Cooperation Program in High Energy Physics (``Negative Ion Drift TPC Development for High-Resolution Tracking'' and ``Research and Development for Current and Future Long Baseline Neutrino Detectors''), as well as the National Science Foundation (EAGER PHY-1649966), and the Research Corporation Cottrell College Science Award (\#23325).


\end{document}